# Liquid-Crystal Smectic with Six-Layer Periodic Structure

P. V. Dolganov[1], V. M. Zhilin[1], V. K. Dolganov[1], and E. I. Kats[2]

[1] *Institute of Solid State Physics, Russian Academy of Sciences, 142432, Chernogolovka, Moscow district, Russia*
[2] *Laue-Langevin Institute, F-38042, Grenoble, France and L. D. Landau Institute for Theoretical Physics, Russian Academy of Sciences, 117940 GSP-1, Moscow, Russia*

Recently Wang *et al.* [Phys. Rev. Lett. **104**, 027801 (2010)] discovered a new polar smectic phase with six-layer period (Sm$C^*_{d6}$). In this manuscript, we demonstrate that the theory [Phys. Rev. E **67**, 041716 (2003)] based on the free energy expansion with a two-component order parameter predicted the six-layer phase, describes its structure and the phase sequence observed in experiment [Phys. Rev. Lett. **104**, 027801 (2010)].

In a recent very interesting paper Wang *et al.* [1] report the discovery of a new polar smectic phase with six-layer period (Sm$C^*_{d6}$). They point that Dolganov *et al.* [2] predict a phase having six-layer period but layer spacing variation among different layers following from [2] has not been observed. The authors [1] made this conclusion on the base of the measurements of resonant X-ray scattering.

In this work, we demonstrate that the theory [2] which is based on the minimization of the free energy with respect to the two-component order parameter predicted the discovered six-layer structure [1]. The theory can describe all Sm$C^*$ type phases observed in the experiments including the six-layer phase (Sm$C^*_{d6}$) and the phase sequence observed in the experiment [1]. We point also [2] that spacing modulation among different layers can be determined by the measurement of the non-resonant X-ray peaks.

For description of Sm$C^*$ variant phases we use the discrete phenomenological Landau model of phase transitions [2] with the two-component order parameters $\xi_i$, where $i$ stands for the $i$th layer. Modulus of the vector $\xi_i$ [Fig. 1(a,b)] is the projection of the long molecular axis onto the smectic plane and characterizes the tilt angle $\theta_i$. Direction of $\xi_i$ describes the angle of azimutal orientation $\varphi_i$ which is the phase of the order parameter. The expansion of the free energy [2-5] is taken in the form $F=F_1+F_2+F_3$, where

$$F_1 = \sum_i \left[ \frac{1}{2} a_0 \xi_i^2 + \frac{1}{4} b_0 \xi_i^4 + f(\xi_i \times \xi_{i+1})_z \right], \quad (1)$$

$$F_2 = \sum_i \left[ a_1 \xi_i \xi_{i+1} + b\xi_i^2(\xi_{i-1}\xi_i + \xi_i \xi_{i+1}) \right], \quad (2)$$

$$F_3 = \sum_i \left[ \sum_{j=2}^{3} a_j \xi_i \xi_{i+j} + a_4 (\xi_i \times \xi_{i+1})^2 \right]. \quad (3)$$

As it is always the case in the phenomenological Landau approach the correspondence in temperature dependent phase sequences between experiment (for a particular material) and theory can be achieved by an appropriate choice of the phenomenological coefficients. In (1), $a_0=\alpha(T-T^*)$ and $b_0$ are the conventional Landau coefficients. $F_2$ and two $a_j$ terms describe interlayer interactions. The third term in (1) is the chiral interaction. The last term in (3) describes an energetic barrier between synclinic ($\varphi_i$-$\varphi_{i+1}$=0) and anticlinic ($\varphi_i$-$\varphi_{i+1}$=$\pi$) ordering in adjacent layers. The physical origin of different terms has been discussed previously [2-5]. The minimization of the free energy was made with respect to the phase and modulus of the order parameter. The method of numerical minimization was described earlier [2].

The structures of all the experimentally observed phases [1] including the Sm$C^*_{d6}$ phase and the sequence Sm$C^*$-Sm$C^*_{d4}$-Sm$C^*_{d6}$-Sm$C^*_\alpha$ [6] corresponding to measurements [1] were obtained in our calculations [Fig. 1(c)] by choosing appropriate values of the model parameters.

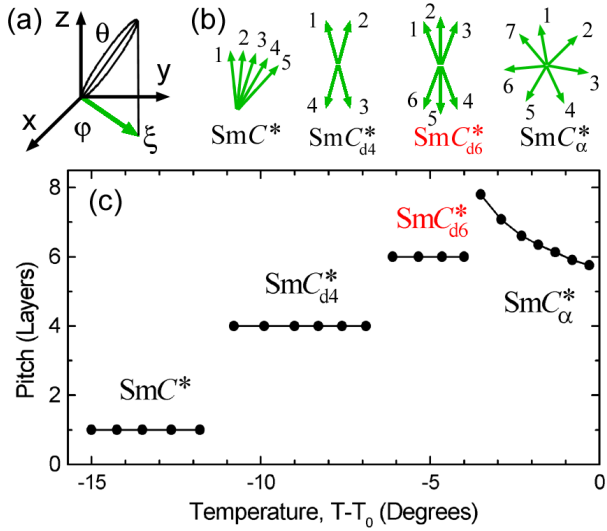

FIG. 1 (a) Schematic representation of the molecular tilt in a smectic layer. (b) Intermolecular orientation structures in different phases [6]. Arrows are the projections of the molecules onto the layer plane. Numbers denote subsequent layers. (c) The phase sequence and temperature dependence of the pitch in different phases obtained in the calculation. The set of parameters is: $\alpha=10^{-2}$ K$^{-1}$, $b_0=1$, $f=3\cdot10^{-4}$, $a_1=-4.68\cdot10^{-3}$, $b=-10^{-3}$, $a_2=2.5\cdot10^{-3}$, $a_3=-10^{-4}$, $a_4=5\cdot10^{-2}$. $T_0$ is the temperature of the transition from the untilted Sm$A$ to the tilted phase.

Layer spacing variation should result in split X-ray peaks centered at wave vectors $Q=Q_0(1+n/6)$, where $n$ is an integer and $Q_0=2\pi/d$, $d$ is the average layer thickness in the six-layer unit cell. The satellite peak at $Q=7/6Q_0$ may be observed only in resonant X-ray scattering. The same situation was observed in experiment [1]. Layer spacing variations should result in a non-resonant peak at $Q=4/3Q_0$ in the Sm$C^*_{d6}$ and Sm$C^*_{d3}$ phases [2,7,8]. This non-resonant peak was indeed found in the Sm$C^*_{d3}$ (Sm$C^*_{FI1}$) phase [9].

In conclusion, the existing theory which is based on the minimization of the free energy with respect to modulus and phase of the order parameter predicted the six-layer structure [2], describes the structure of polar phases including the Sm$C^*_{d6}$ phase and the phase sequence observed in experiments [1].

This work was supported in part by RFFI Grant 08-02-00827.